\documentclass[10pt]{iopart}
\usepackage{graphicx}
\usepackage{float}

\begin{document}

\title{On spontaneous fission and $\alpha$-decay half-lives of atomic nuclei}

\author{K Pomorski, M Warda, A Zdeb}

\address{Department of Theoretical Physics, Maria Curie-Sk{\l}odowska
    University, Lublin, Poland}
\ead{azdeb@kft.umcs.lublin.pl}
\vspace{10pt}
\begin{indented}
\item[]November 2015
\end{indented}

\begin{abstract}
It is shown, that the Gamow-like model with only one adjustable parameter --
radius constant is able to reproduce well the alpha-decay half-lives
for all even-even nuclei with the proton number larger than 50. The
systematics for odd-A and odd-odd isotopes can be also well described when
ones introduces an additional hindrance factor. Similar model based on the W.
J. \'Swi\c{a}tecki idea from 1955 is developed to reproduce the spontaneous
fission half-lives of transactinide nuclei. The achieved accuracy of
reproduction of the data is better than that obtained in more advanced
theories.
\end{abstract}


\vspace{2pc}

\submitto{\PS}
%
\maketitle
%
\ioptwocol

\section{Introduction}

Large progress in synthesis of heavy nuclei done in the last decades raises new
demands for a better and better theoretical description of their decay modes.
Spontaneous fission and $\alpha$-radioactivity are the most important processes
of disintegration of heavy nuclei.

Alpha decay occurs most often in actinides region, but is also observed in
isotopes with $Z\ge 52$. The first theoretical interpretation of alpha-decay
process was given independently by Gamow \cite{G28}, Gurney and Condon
\cite{GC28} in 1928 year. Emission of alpha particle is treated as
quantum-mechanical tunnelling through the nuclear Coulomb barrier, where the
probability of emission is calculated using one-dimensional WKB approximation.
Recently it was proven, that this approximation can be successfully used to
evaluate the probability of tunnelling by alpha particle and cluster as well
(see Ref.~\cite{ZWP13}). It was also shown, that within this simple model
(containing only 1 adjustable parameter for even-even nuclei) one can reproduce
alpha decay half-lives of heavy emitters with higher accuracy in comparison
with modern (containing 5 parameters) version~\cite{PS05} of Viola-Seaborg
formula~\cite{VS66}.

The observation in 1938 of the neutron induced nuclear fission by Hahn and
Strassmann came rather unexpected \cite{HS39}. This new phenomenon was explained
within a few weeks by Meitner and Frisch's who established the most
important features of low-energy fission: the energy released in this process is
equal almost 200 MeV and results from the Coulomb repulsion of the fission
fragments and the number of neutron emitted per fission event larger than one,
what has opened a possibility for a chain reaction \cite{MF39}. One and a half
year after Hahn and Strassmann's discovery, Flerov and Petrzhak have first
observed the spontaneous fission of Uranium \cite{FP40}. As we have written
above, Gamow had explained the $\alpha$-decay and its sometimes rather long
half-lives by a quantum tunnelling process of a preformed $\alpha$ particle
through a Coulomb barrier. So, according to the concepts of Meitner and Frisch,
one could expect spontaneous fission of uranium also from the ground-state, but
with a considerably longer half-life than for the $\alpha$-decay, because of the
larger reduced mass for almost symmetric fission.

The first quantitative estimates of the spontaneous fission probability becomes
possible, when the microscopic-macroscopic model of the potential energy of
deformed nuclei was developed and the Inglis cranking model was implemented to
evaluate the inertia corresponding to the fission mode. Of course in time these
models give more and more precise reproduction of the spontaneous fission
half-lives systematics and the fission mass distributions. A short critical
description of the spontaneous fission theories will be presented in our paper
in order to better understand physical background of a simple model for the
probability of this decay mode, which we have developed following an old idea of
W. J. \'Swi\c{a}tecki. Namely, in 1955 year he proposed a formula which was
able to describe the global systematics of the spontaneous fission
half-lives~\cite{Sw55}. His simple phenomenological formula, based on
correlations between logarithms of observed spontaneous fission half-lives and
ground state microscopic corrections, reproduced well the experimental data
known at that time. We are going to show in the following that his idea,
combined with the modern version of the liquid drop model (LSD)~\cite{PD03},
allows to obtain satisfactory agreement with the measured up to now spontaneous
fission half-lives.

The paper is organized as follows: the main assumptions of $\alpha$-decay model
and results obtained for emitters with $52\le Z\le 110$ are presented in Sec. 2.
The WKB theory of the fission barrier penetration as well as the the
semi-empirical \'Swi\c{a}tecki's formula for spontaneous fission half-lives will
be described in Sec. 3, where results for the isotopes with $90\le Z\le 114$ are
analyzed. Sec. 4. contains Summary.

\section{Alpha decay}%

\subsection{The model}

The quantum tunnelling theory of alpha emission assumes, that the decay
constant $\lambda$ is proportional to the barrier penetration probability $P$,
frequency of assaults on the nuclear Coulomb barrier per time-unit $n$ and
particle preformation factor
$S_{\alpha}$. In the presented model the expression for decay constant is
simplified as the effect of the preformation is included into the probability
$P$ (see discussion in Ref.~\cite{ZWP13}):
\begin{equation}
\lambda=n P\,\,.
\label{lambda}
\end{equation}
The probability $P$ of tunnelling through the barrier is calculated using
one-dimensional WKB approximation:
\begin{equation}
P=\exp{\displaystyle\biggl[-\frac{2}{\hbar}\int_{R}^{b}
 \sqrt{2\mu(V(r) -E_{\rm\alpha})}dr\biggr]}\,\,,
\label{P}
\end{equation}
%
%
\begin{figure}[H]
\includegraphics[width=1\columnwidth]{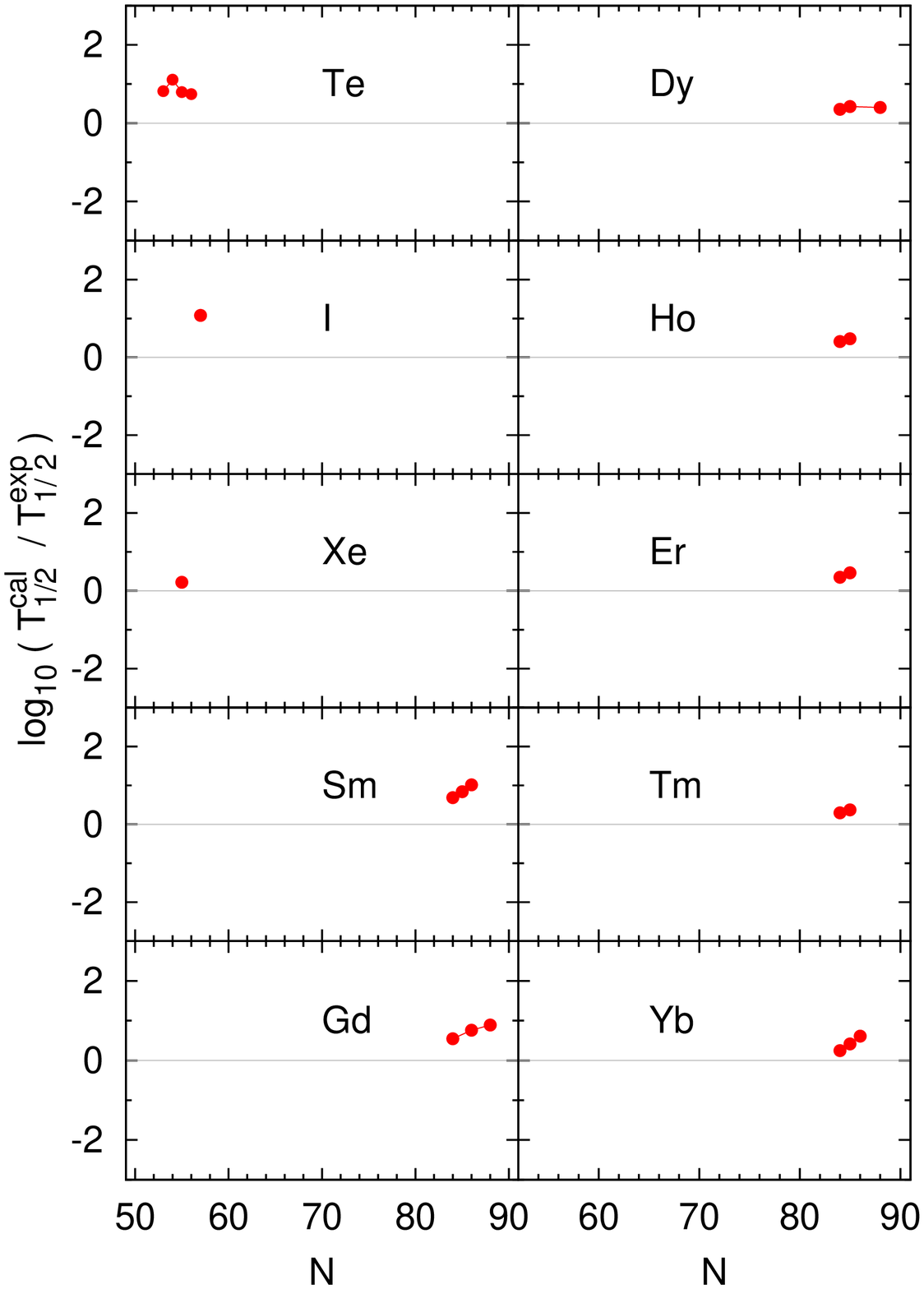}
\includegraphics[width=1\columnwidth]{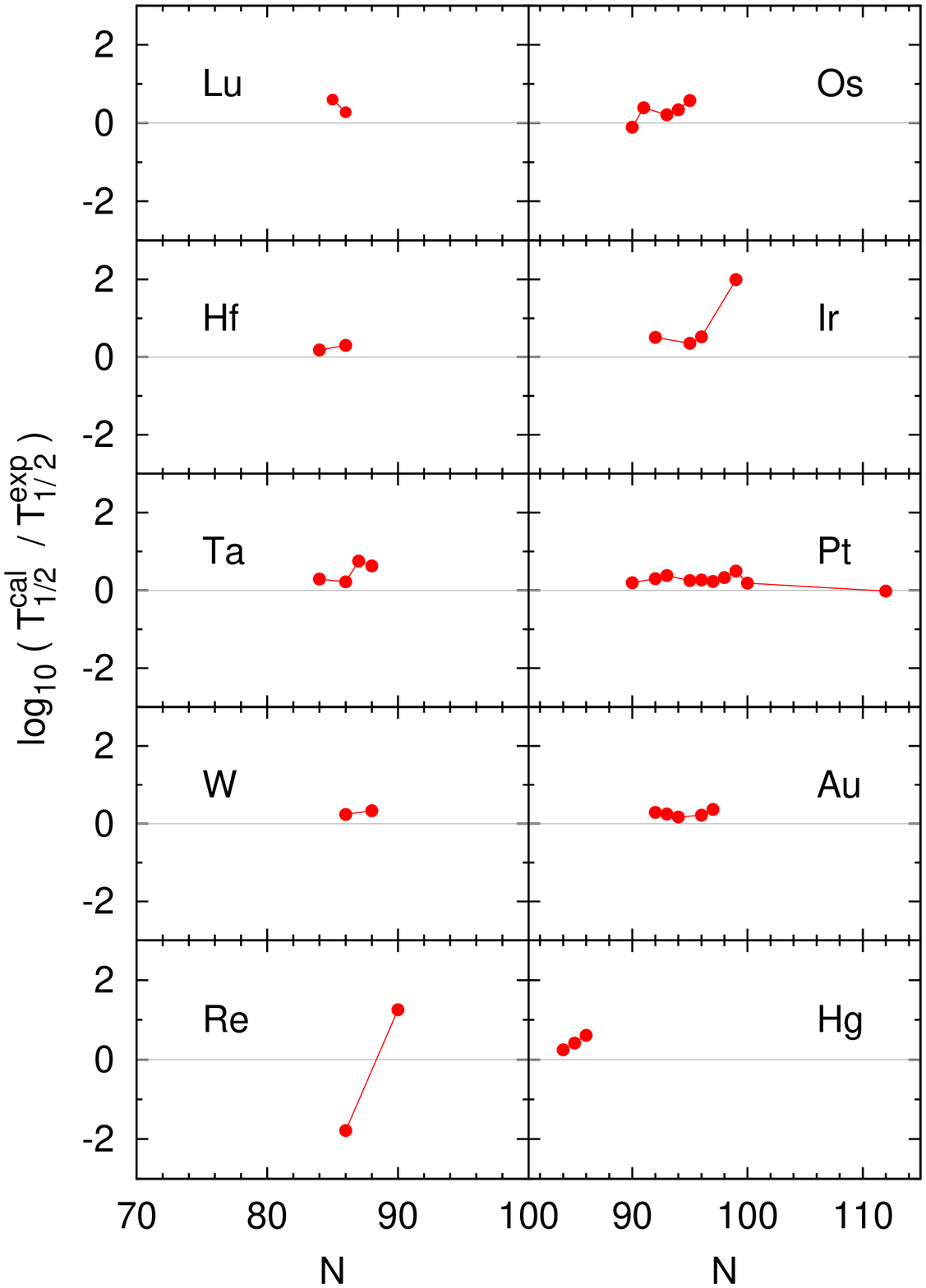}
\caption{Differences of the theoretical and experimental~\cite{nudat2} alpha
decay half-lives on logarithmic scale, calculated for nuclei with $52 \le Z \le
70$ (top) and $71 \le Z \le 80$ (bottom). }
\label{fig1}
\end{figure}
\begin{figure}[H]
\includegraphics[width=1\columnwidth]{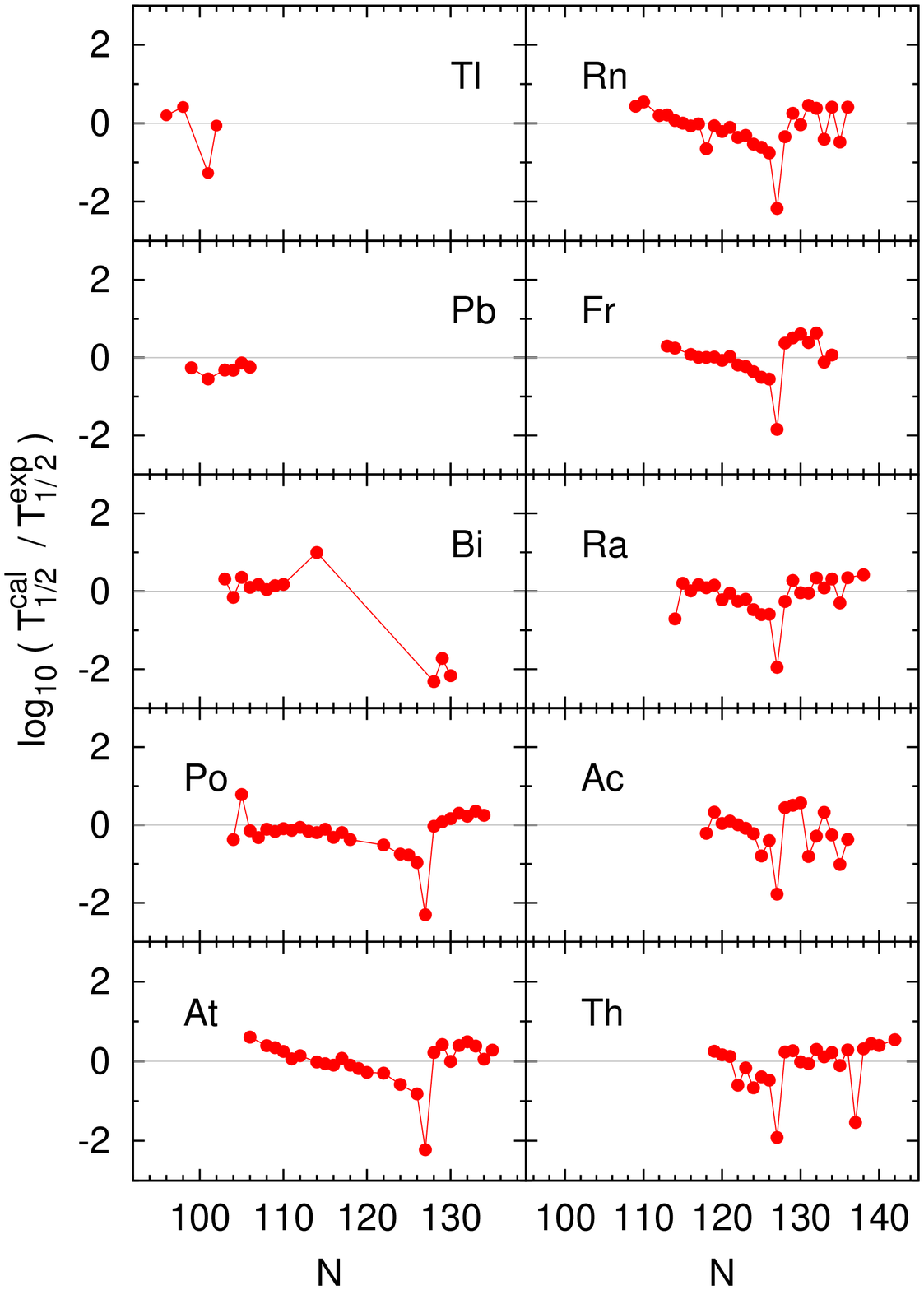}
\includegraphics[width=1\columnwidth]{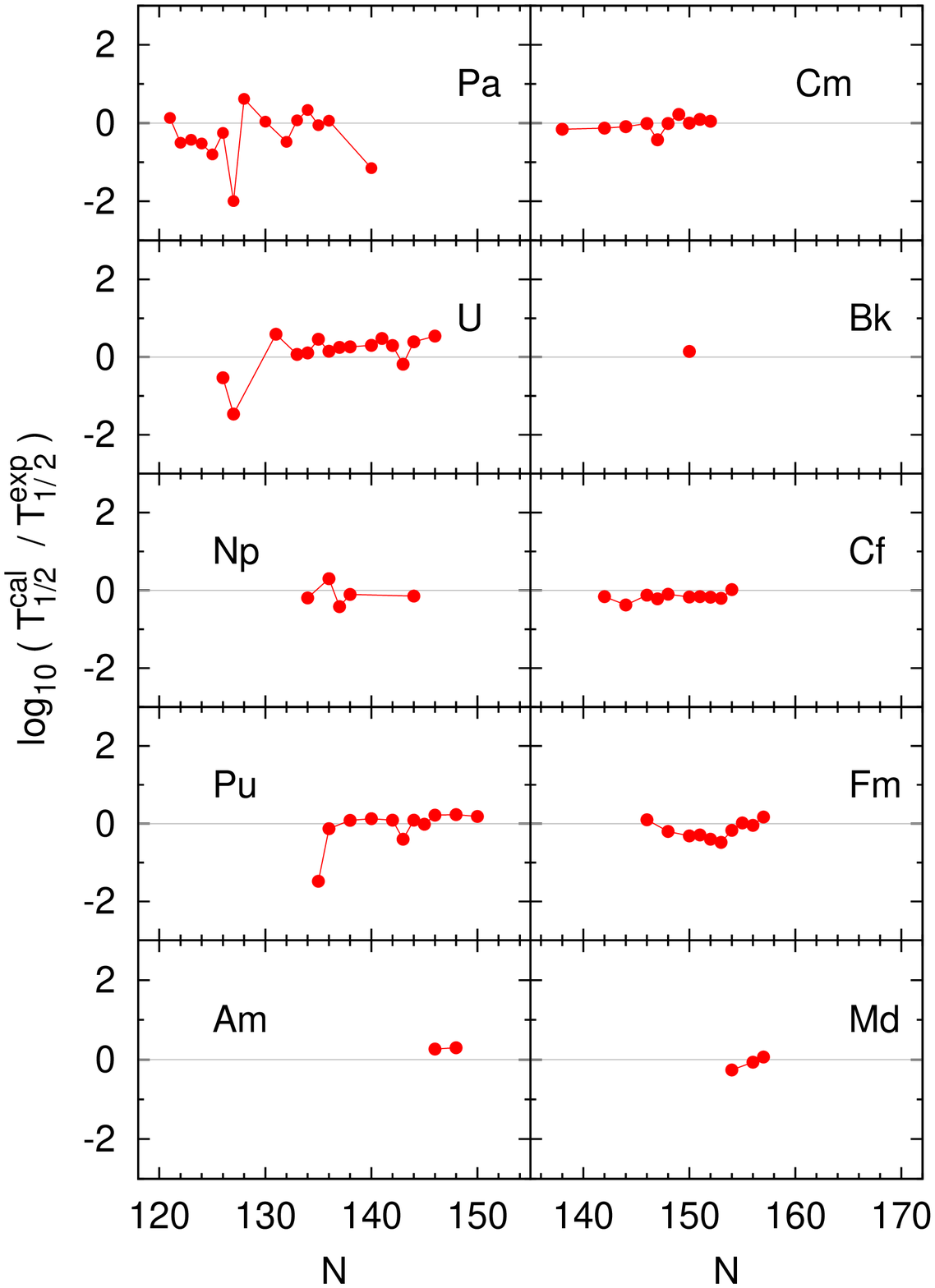}
\caption{The same as in Fig. \ref{fig1}, but for isotopes with $81 \le Z \le 90$
(top) and $91 \le Z \le 101$ (bottom). } \label{fig2}
\end{figure}
\begin{figure}[H]
\includegraphics[width=1\columnwidth]{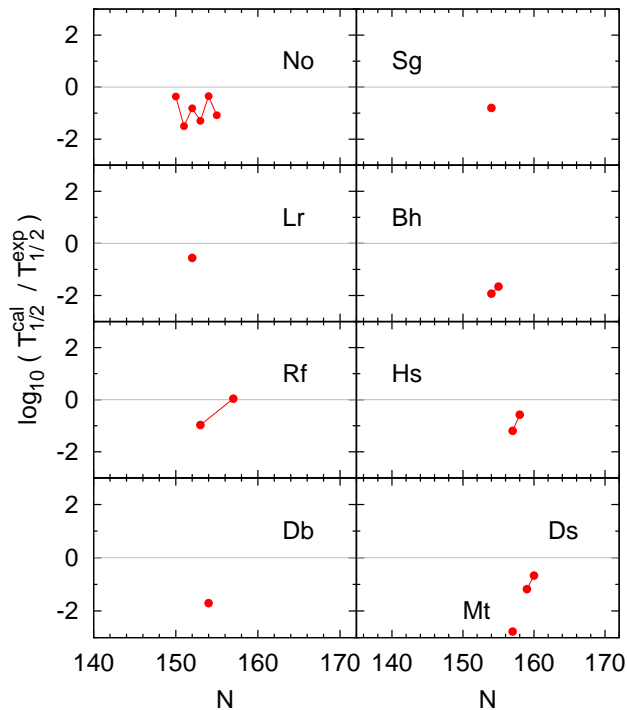}
\caption{The same as in Fig. \ref{fig1}, but for isotopes with $102 \le Z \le
110$.}
\label{fig3}
\end{figure}

\noindent
where $\mu$ is a reduced mass, $R$ is the spherical square well radius
\begin{equation}
R=r_{0}(A_{\alpha}^{1/3}+A_{d}^{1/3})
\end{equation}
and $b$ is the exit point from the Coulomb barrier:
\begin{equation}
b=\frac{Z_{\alpha}Z_{d}e^{2}}{E_{\alpha}}~.
\end{equation}
$A_{\alpha},Z_{\alpha}$ and $A_{d},Z_d$ are the mass and proton numbers of an
alpha particle (or cluster) and a daughter nucleus respectively. $E_{\alpha}$ is
the kinetic energy of emitted alpha (cluster) particle \cite{nudat2}. The number
of assaults per time-unit $n$ is evaluated from the quantum-mechanical
ground-state frequency in the spherical square well:
\begin{equation}
n=\frac{\pi\hbar}{2\mu R^2}\,\,.
\end{equation}
In this formalism the $\alpha$-decay half-life can be expressed as follows:
\begin{equation}
T_{1/2}=\frac{\ln 2}{\lambda} 10^h\,\,,
\label{Talfa}
\end{equation}
where constant $h$ (so-called {\it hindrance factor}) was additionally
introduced for odd nuclei. The least-square fit of the radius constant $r_0$ was
performed to known experimental half-lives of even-even $\alpha$-emitters (127
cases)~\cite{nudat2}. Only the most probable alpha decays mode at each isotope
was chosen in our analysis. Obtained in this way value of the
nuclear well radius constant is equal to $r_0= 1.23$~fm, and slightly differs from
that reported in Ref.~\cite{ZWP13}, as now more alpha emitters were taken and
the cluster decays were not included in our analysis. Fitting procedures of
{\it hindrance factor} were performed for odd systems with fixed $r_0$ value
($h=0.25$ for odd-A and doubled for odd-odd emitters).

\subsection{Results}

The logarithms of the ratios of the half-lives ($T^{\rm cal}_{1/2}$),
calculated using above formalism, to the measured ones ($T^{\rm exp}_{1/2}$)
for all examined nuclei are shown in Figs.~\ref{fig1}-\ref{fig3} as a function
of the neutron number $N$.

Large deviations from the data are observed for emitters with neutron number
$N=127$ in $_{83}$Bi - $_{92}$U elements (Fig.~\ref{fig2}), where
underestimations of the $\rm log_{10}(T_{1/2})$ reach about two orders of
magnitude. Calculated half-lives of some lighter emitters ($_{52}$Te - $_{74}$W)
are overestimated, but these discrepancies do not exceed one order of magnitude.
The root-mean-square deviations of our estimations made for the $\alpha$-decay
half-lives for all considered nuclei are summarized in Table 1.

\begin{center}
\begin{table}[H]
\caption{Root-mean-square deviations of $\log_{10}(T_{1/2}^{\alpha})$
calculated using Eq. (\ref{Talfa}) with the radius constant $r_0$=1.23 fm.\\}
\begin{center}
\begin{tabular}{crlc}
\hline
$\pi_Z-\pi_N$&~~~n~~~&~~~~~h~~~~~ &~~r.m.s.~~\\[0.5ex]
\hline
   e-e   & 127& ~~~~0  & 0.39 \\
   e-o   & 97 & ~~~~0.25 & 0.66 \\
   o-e   & 82 & ~~~~0.25 & 0.54 \\
   o-o   & 54 & ~~~~0.5 & 0.79 \\[0.5ex]
\hline

\end{tabular}
\end{center}
\end{table}
\end{center}

\section{Spontaneous fission within the WKB theory}
\label{SF}

A continuous interest in the theoretical description of the fission dynamics is
observed since discovery of this phenomenon in 1938. In present Section we are
going to concentrate on the spontaneous fission and its description within the
WKB theory only. The results presented here will be used to understand, why a
simple one parameter (for even-even nuclei) model {\it \`a la \'Swi\c{a}tecki}
\cite{Sw55}, described in Section~\ref{alaSw}, is able to reproduce the
spontaneous fission  half-lives of all known nuclei with higher accuracy than
obtained using more advanced theories.

Quantum mechanically it is possible for a nucleus in its ground state to tunnel
the fission barrier. The probability of the barrier penetration depends not only
on the height and width of the barrier but also on the magnitude of the
collective inertia associated with the fission mode. The calculations of the
spontaneous fission half-lives $T_{\rm sf}$ require a careful evaluation of the
collective potential energy surface and the collective inertia tensor. Different
models, like the macroscopic-microscopic model \cite{NTS69,BDJPSW} or the
Hartree-Fock-Bogolubov self-consistent theories (see e.g. \cite{RP75}), can be
used to generate the potential energy surfaces. The collective mass parameters
one evaluates usually in the cranking approximation \cite{SSW69,BDJPSW} or
within the Generator Coordinate Method (GCM) with the generalized Gaussian
Overlap Approximation (GOA) \cite{GPB85}.
\begin{figure}[h]
\centerline{\includegraphics[width=8cm]{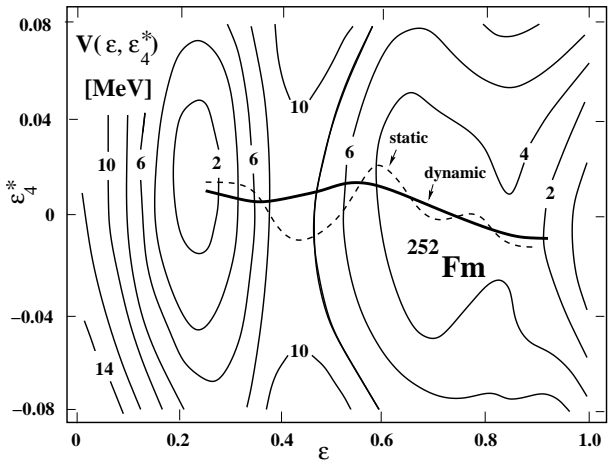}}
\caption{A contour plot of the energy surface for $^{252}$Fm is shown in the
two-dimensional quadrupole $(\epsilon)$ and hexadecapole $(\epsilon_4)$
deformation parameter space. The dashed line is the path of steepest descent
from the saddle points to the ground state and to the scission line, the thick
 line is the minimal-action path. After Ref.~\cite{Bar78}.}
\label{Baran}
\end{figure}

The spontaneous fission half--life is given by:
\begin{equation}
  T^{\rm sf}_{1/2}=\frac{\ln2}{nP}\,\,.
\label{tsf}
\end{equation}
Here $n$ is the number of assaults per time unit, associated with the
frequency of zero--point vibration of the nucleus in the fission mode direction.
The fission barrier penetration probability $P$ might be evaluated within the
one-dimensional WKB approximation, what leads to the following expression:
\begin{equation}
 P = \frac{1}{1+\exp\{2S(L)\}} \,\,.
\label{WKB}
\end{equation}
The action--integral $S(L)$, calculated along a fission path $L(s)$ in the
multi--dimensional space of collective coordinates is given by:
\begin{equation}
 S(L) = \int^{s_2}_{s_1} \left\{{2 \over \hbar^2} \, B_{\rm eff}(s)
 [V(s) - E_{\rm gs}]\right\}^{1/2} ds \,\,,
\label{act}
\end{equation}
where the integration limits $s_1$ and $s_2$, correspond to the classical 
turning points. $E_{\rm gs}$ is the energy of a nucleus in its ground state and $V(s)$ is
the collective potential. $B_{\rm ss}(s)$ is an effective inertia tensor in the
multi-dimensional space of collective coordinates $\{q_i\}$:
\begin{equation}
 B_{\rm ss}(s) = \sum_{k,\,l} \, B_{kl} \,{dq_k \over ds} {dq_l \over ds}.
\label{beff}
\end{equation}

\begin{figure}[htb]
\centerline{\includegraphics[width=8cm]{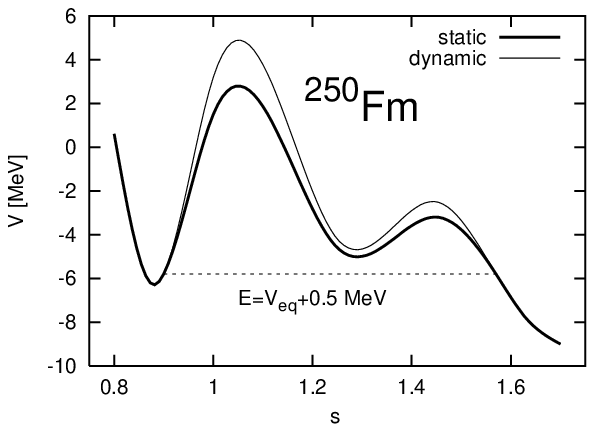}}
\centerline{\includegraphics[width=8cm]{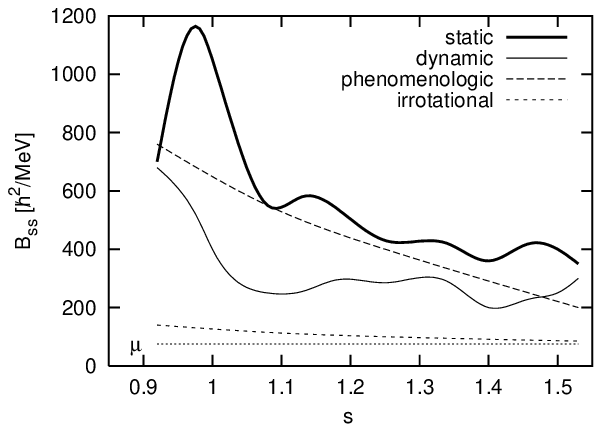}}
\caption{Collective potential (top) and inertia $B_{\rm eff}$ (bottom) along
the static (thick solid line) and the dynamic (thin solid line) path to fission
as a function of the relative distance between the fragment mass centres
$s=R_{12}/R_0$. The
reduced ($\mu$), irrotational flow and phenomenological inertias
\protect\cite{Ran76} are shown for comparison. After Ref.~\cite{KP08}.}
\label{fig5.1}
\end{figure}

\begin{figure}[htb]
\centerline{\includegraphics[width=8cm,angle=0]{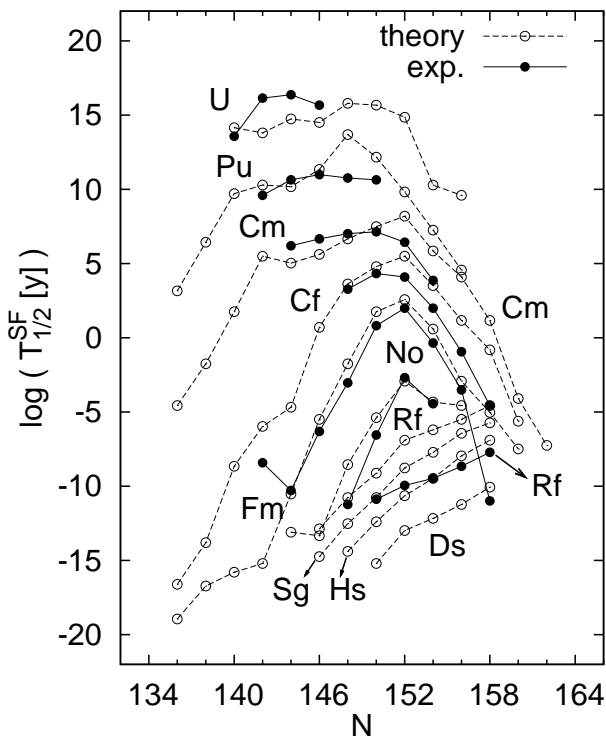}}
\caption{Spontaneous fission half-lives of even-even transactinides and their
shape isomers (II) obtained with the WKB approximation (\ref{WKB}) along the
least action (dynamical trajectory). After Ref.~\cite{Ba81}.}
\label{dynamic}
\end{figure}

Determination of the fission path in the deformation space plays crucial
role during evaluation of the fission probability~\cite{BDJPSW}. The action
integral (\ref{act}) should be calculated along such a path (so-called {\it
dynamic}), which minimize its total value, thereby maximize fission probability
(\ref{WKB}) \cite{Ba81}. In contrast to the {\it static path}, the dynamic one (dependent on
the inertia tensor) does not have to lead through the bottom of the fission
valley, as one can see in Fig.~\ref{Baran}. The comparison of the fission
barrier and effective inertia, corresponding to the dynamic and static fission
paths for $^{250}$Fm is shown in Fig.~\ref{fig5.1}. The
collective potential $V(s)$, corresponding to the dynamic path, is larger than
in case of the static one, but the dynamic inertia is smaller. It might be
roughly approximated by the so-called {\it phenomenological inertia} fitted to
the observed spontaneous fission half-lives~\cite{Ran76}:
\begin{equation}
{\cal M}^{\rm phen}_{R_{12}R_{12}} = \mu\left(1+k\frac{17}{15}
    \exp\left[-a(R_{12}-\frac{3}{4})\right]\right) \,\,.
\label{Mphen}
\end{equation}
Here $\mu$ is the reduced mass, corresponding to the relative distance between
fragments ($R_{12}$=3/4 for a sphere) and $a=2.452$ is a numerical constant,
determined from a fit to the exact irrotational flow inertia \cite{FN72} for
$k=1$. It was found in Ref.~\cite{Ran76} that the phenomenological inertia
obtained with the constant $k=11.5$ reproduces well the systematics of the known
spontaneous fission half-lives.

Spontaneous fission half-lives of even-even transactinides evaluated in
Ref.~\cite{Ba81} are presented in Fig.~\ref{dynamic}. The estimates correspond
to the least-action trajectories (dynamic path to fission). The WKB
approximation (\ref{WKB}) and the cranking inertia tensor were used. The
two-dimensional potential energy surfaces for each isotope were obtained within
the macroscopic-microscopic model using the Nilsson potential and the droplet
energy for the macroscopic part. The effects of the axial asymmetry as well as
the reflection asymmetry were taken into account.

Usually calculations of spontaneous fission half-lives are performed with an
assumption, that the collective potential is equal to the Hartree-Fock-Bogolubov
(HFB) binding energy or its macroscopic-microscopic approximation. The
zero-point energy corrections are often neglected in papers dealing with similar
problems. We shell discuss this point later. Moreover, one assumes, that the
ground state is located at $\Delta E\approx 0.5$ MeV above the minimum of the
HFB potential. As a justification of such a choice of $\Delta E$ one refers to
experimental values of the average quadrupole phonon energy, which is about 1 MeV for
actinides. Several authors even treat $\Delta E$  as a free parameter. It should
be stressed, that the HFB theory, as each variational method, gives the upper
limit of the ground-state energy, so the zero-point correction energy at the
equilibrium point is approximately equal to the difference of the ground-state
and the potential energy at the minimum (see Ref.~\cite{Goe77}).  It can be
easily shown, that the zero-point energy of harmonic oscillations is equal to
the half of the corresponding phonon energy, i.e. ($\frac{1}{2}\hbar\omega$)
\cite{RS80}, what is consistent with result, obtained in Ref.~\cite{Rob88},
where the coupled quadrupole and octupole vibrations in the region of Ra--Th
nuclei were analyzed. This conclusion is also valid even for strongly anharmonic
motion such as pairing vibrations~\cite{Sie04}.

The number ($n$) of the collective degrees of freedom results much on the
value of the zero-point energy correction. Namely, the $\varepsilon_0$ is
proportional to $n\cdot\hbar\omega$. Thus, when the HFB equations are solved in the
whole $n$-dimensional space of collective coordinates, all zero-point
corrections should be substracted from the energy to obtain the collective
potential:
\begin{equation}
 V(s) = E_{\rm HFB}(s) - \varepsilon_0^{\rm fiss}(s) \,\,,
\label{ezerof}
\end{equation}
where $\varepsilon_0^{\rm fiss}$ is the zero-point energy related to the fission
mode \cite{GPB85}.

In the case of the one dimensional fission barrier penetration given by
the Eq. (\ref{act}), the oscillations perpendicular to the fission path are not
included during evaluation of tunnelling probability~(\ref{Tsf}). In this
direction the zero-point energy corrections will be approximately cancelled by
the 1/2 of the corresponding phonon energy.

\begin{figure}[htb]
\centerline{\includegraphics[width=8cm]{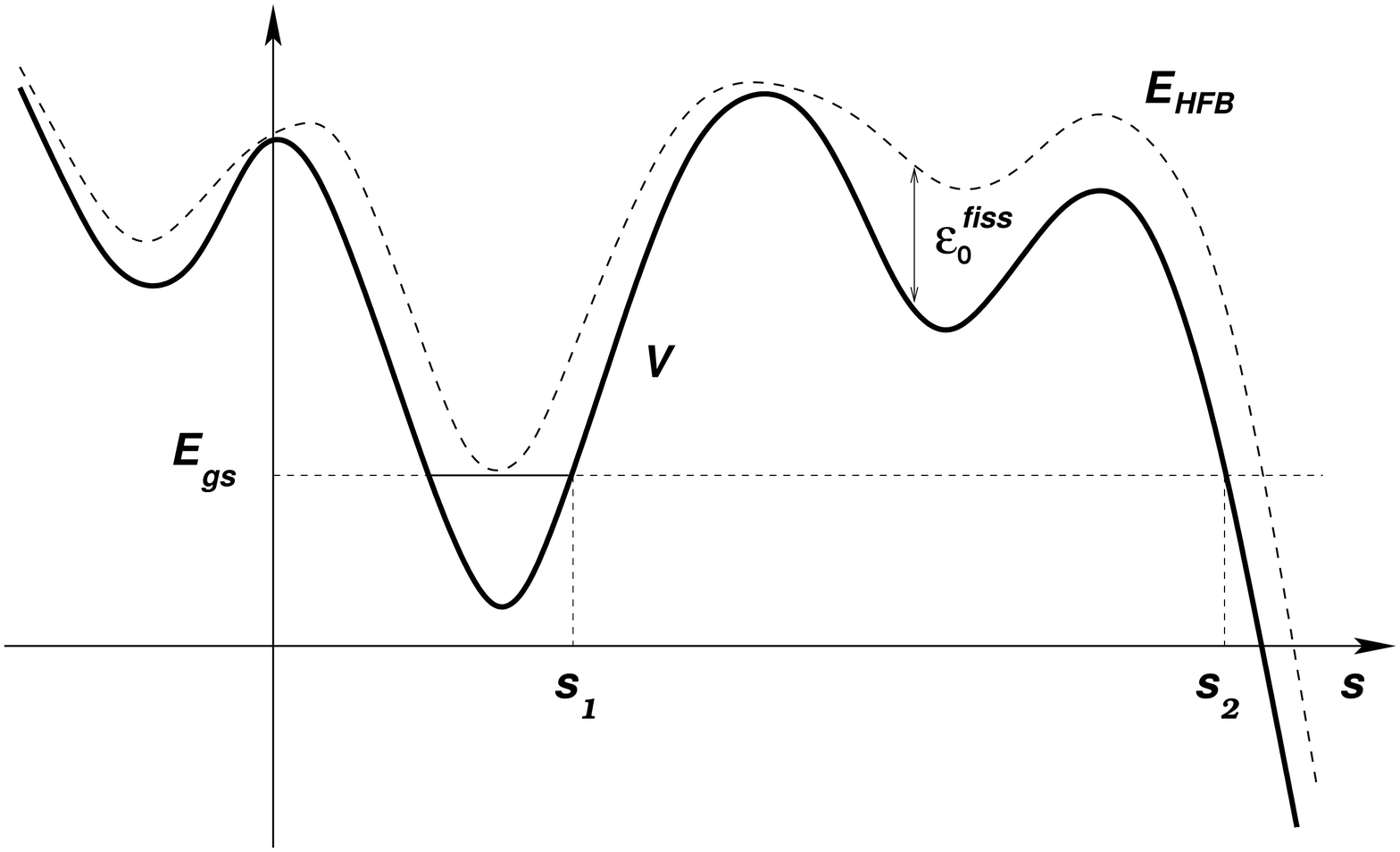}}
\caption{Collective potential $V(s)$, binding energy $ E_{\rm HFB}(s)$ and
ground-state energy $E_{\rm gs}$ of a nucleus along the fission path $s$. After
Ref.~\cite{KP08}.}
\label{fig5.2}
\end{figure}

In Fig.~\ref{fig5.2} the potential and the binding energy along the fission path
are plotted. As one can observe, the zero-point corrections may affect on the
fission barrier height.

The determination of the least action path in deformation space is not a trivial
task (see Ref.~\cite{BDJPSW} for further discussion). The approach, which is
worth to be mentioned in this context is dynamical programming method
~\cite{Bar78,Ba81}. Nevertheless, applying this method one should choose optimal
distance between the mesh points to avoid large numerical errors. That is
because the accuracy of calculations of the partial derivatives $\frac{\partial
q_i}{\partial s}$, which result much on the effective mass parameter
(\ref{beff}) as well as the zero-point energy correction (see e.g.
Ref.~\cite{GPB85}) along the fission path, strongly depends on the grid size. 
The von Ritz implementation (see appendix of Ref.~\cite{DPB07}) might be useful
as an alternative for practical calculations of the spontaneous fission
probability. The main advantage of this method is that the influence of the
zero-point energy correction along the dynamical paths might easy be taken into
account.

Typically the mass yield of the spontaneous fission fragments of the actinide
nuclei is spread between 70 and 180. Usually, except the bimodal fission of
$^{258}$Fm and neighbouring nuclei, the asymmetric fission is more probable
than the symmetric one and the most populated mass of the heavier fragment is
around 140. A very mass asymmetric spontaneous fission process, when the mass of the lighter fragment is around 20, is already considered as a cluster emission. This type of radioactivity was predicted in 1980 by Sandulescu and co-workers \cite{SPG80} and four years later it was discovered by Rose and Jones
\cite{RJ84}, who observed the spontaneous emission of $^{14}$C from
$^{223}$Ra. The cluster emission is a very rare process. Its relative branching ratio to the $\alpha$-decay is of the order $10^{-10}$ to $10^{-17}$. Nevertheless, in last two decades one has observed clusters from $^{14}$C to
$^{34}$Si emitted by actinide nuclei from $^{221}$Fr to $^{242}$Cm. In all
cases the residual nucleus was always close to the double magic $^{208}$Pb.
\begin{figure}[htb]
\centerline{\includegraphics[width=\columnwidth]{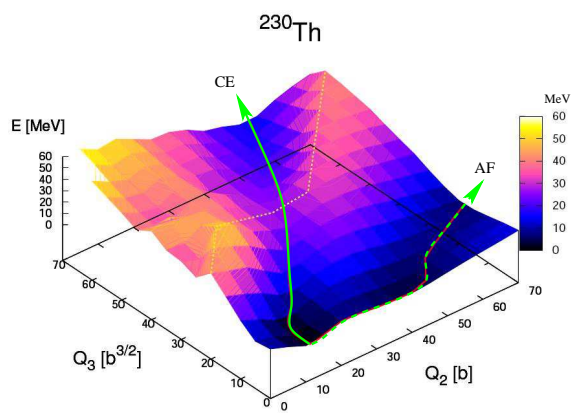}}
\caption{The HFB potential energy of $^{230}$Th as a function of quadrupole and
octupole moments. The the thick solid line show the path to cluster emission
(CE) while the dashed one is drawn along the valley to asymmetric fission (AF). After Ref.~\cite{RWa08}.}
\label{warda1}
\end{figure}

In general, there are two alternative theoretical descriptions of the
spontaneous cluster decay. One can use a fission-like mechanism to reproduce
the main features of the process \cite{SPG80,PP86,SSw87,PGG06,RWa08} or one
assumes, in close analogy to the $\alpha$-decay, that a cluster, performed by a
nonadiabatic mechanism inside nucleus, penetrates the potential barrier
created by the Coulomb and nuclear interaction with a daughter nucleus
\cite{KFT86,GGM87,BFW87,BFW91}. The potential energy landscape of $^{230}$Th
obtained by the HFB calculation with the Gogny D1S force \cite{RWa08} is
presented in Fig.~\ref{warda1} as a function of quadrupole ($Q_2$) and octupole
($Q_3$) moments. Two valleys leading to the cluster emission (CE, solid line)
and to the asymmetric fission (AF, dashed line) are visible in the plot.

The above theoretical models, which describe the cluster-radioactivity, are
rather complex and contain adjustable parameters. Last year we have shown in
Ref.~\cite{ZWP13}, that within a simple Gamow model with only one adjustable
parameter (radius constant) common for the $\alpha$-decay and the cluster
emission one can describe with good accuracy the experimental systematics of
half-lives for the cluster radioactivity of even-even nuclei. An additional
hindrance constant was introduced in Ref.~\cite{ZWP13} to describe the cluster
emission probability from odd-even, even-odd and odd-odd isotopes.

\section{Simple phenomenological formula for the spontaneous fission
half-lives}
\label{alaSw}

Encouraged by the good result for $T_{1/2}^\alpha$ obtained in the Gamow
theory, we are going in the following to describe in another simple model the
spontaneous fission half-lives of transactinide nuclei. Namely, we shall adopt
the \'Swi\c{a}tecki idea from 1955 \cite{Sw55}, who found a simple
relation between the spontaneous fission half-lives and the experimental mass
deviations from their liquid drop estimates. The crucial ingredient of the
model is the liquid drop formula which is described in the next subsection.

\subsection{The liquid drop model}

In our analysis we decided to use the modern version of the
macroscopic-microscopic model~\cite{PD03}. The macroscopic part of this
Lublin-Strasbourg Drop (LSD) mass formula (in MeV units) is as follows:
\begin{equation}
\begin{array}{l}
M_{\rm LSD}(Z, N, {\rm def})=\\[+0.5ex]
~~ 7.289034\cdot Z+8.071431\cdot N-0.00001433\cdot Z^{2.39}\\[+0.5ex]
~ -15.4920(1 - 1.8601 I^2) A\\[+0.5ex]
~ + 16.9707(1 - 2.2938 I^2) A^{2/3}\, B_{\rm surf}({\rm def})\\[+0.5ex]
~ + 3.8602(1 + 2.3764 I^2) A^{1/3}\,B_{\rm cur}({\rm def})\\[+0.5ex]
~ + 0.70978\,Z^2/A^{1/3}\,B_{\rm Coul}({\rm def})
  - 0.9181\,Z^2/A\\[+0.5ex]
~ -10\exp({-4.2|I|})\,B_{\rm cong}({\rm def})+E_{\rm o-e} \,\,.
\end{array}
\label{LSD}
\end{equation}
where the odd-even energy term $E_{\rm e-o}$ is given in Ref.~\cite{MS96}. In
Eq.~(\ref{LSD}) $A=Z+N$ denotes the mass number, $I=(N-Z)/A$ reduced isospin
and $B_{\rm surf}$, $B_{\rm cur}$, $B_{\rm Coul}$ and $B_{\rm cong}$ are
relative to the sphere: surface, curvature, Coulomb and congruence (see
Ref.~\cite{MS96}) energies. The parameters in the first and the last row in
Eq.~(\ref{LSD}) are taken from Ref.~\cite{MN95}, while the rest 8 parameter
were fitted to the data.

\subsection{Simple model \`a la \'Swi\c{a}tecki}

Following the idea presented in Ref.~\cite{Sw55}, we are going to find an approximative functional dependence of the logarithms of spontaneous fission half-lives, corrected by mass shifts $k\delta M$, on proton $Z$ and neutron $N$ numbers:
\begin{equation}
f(Z,N)=\log_{10} [T_{1/2}^{\rm sf}/y]+k\delta M(Z,N)\,\,.
\label{fZN}
\end{equation}
The $\delta M$ is a ground-state microscopic energy, defined as:
\begin{equation}
\delta M_{micr}^{\rm exp}(Z,N)=M_{\rm exp}(Z,N)-M_{\rm LSD}(Z,N,0),
\label{dM}
\end{equation}
where $M_{\rm exp}(Z,N)$ is a measured mass of isotope, taken from~\cite{nudat}
and $M_{\rm LSD}(Z,N,0)$ were calculated using formula (\ref{LSD}).
\begin{figure}[h!]
\begin{centering}
\includegraphics[width=0.75\columnwidth, angle=270]{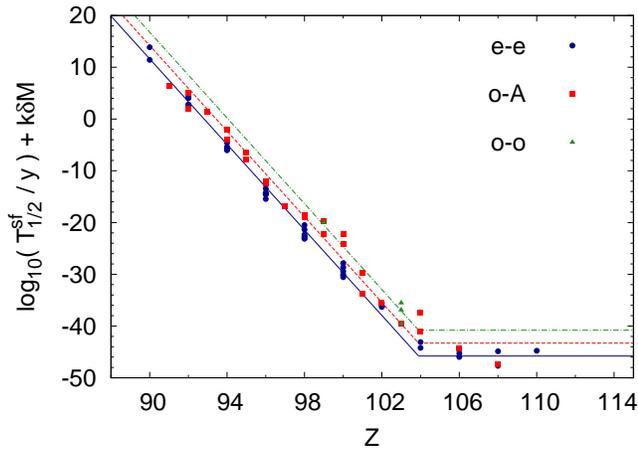}
\caption{Logarithms of the observed spontaneous fission half-lives corrected
with masses "shifts" $k\delta M$ as a function of proton number.}
\label{fit}
\end{centering}
\end{figure}

Fitting procedures included 35 fissioning even-even isotopes ($Z\le 102$) with
measured masses and half-lives. The smooth dependence for even-even isotopes
was achieved for factor $k=7.7\, \rm MeV^{-1}$, as it is shown in
Fig.~\ref{fit}. The curves, fitted for odd-$A$ and odd-odd nuclei are shifted
by a constant- {\it hindrance factor}. It is equal to $h=2.5$ for odd-$A$
isotopes and doubled for odd-odd systems. Formula for spontaneous fission
half-lives in years is given by:
\begin{equation}\label{Tsf}
\begin{array}{l}
\log_{10}\left(\frac{T_{1/2}^{\rm sf}}{y}\right)
  =-4.1\cdot {\rm min}(Z,103)+380.2\\
~~~ -7.7\,\delta M(Z,N)
+\left\{
\begin{array}{ll}
0  &{\rm ~for~even-even \,\,,}\\
2.5  &{\rm ~for~odd-}A\,\,,\\
5&~ {\rm for~odd-odd\,\,.}
\end{array}\right.
\end{array}
\end{equation}

\begin{figure}[ht!]
\hskip0.15cm
\includegraphics[width=0.75\columnwidth, angle=270]{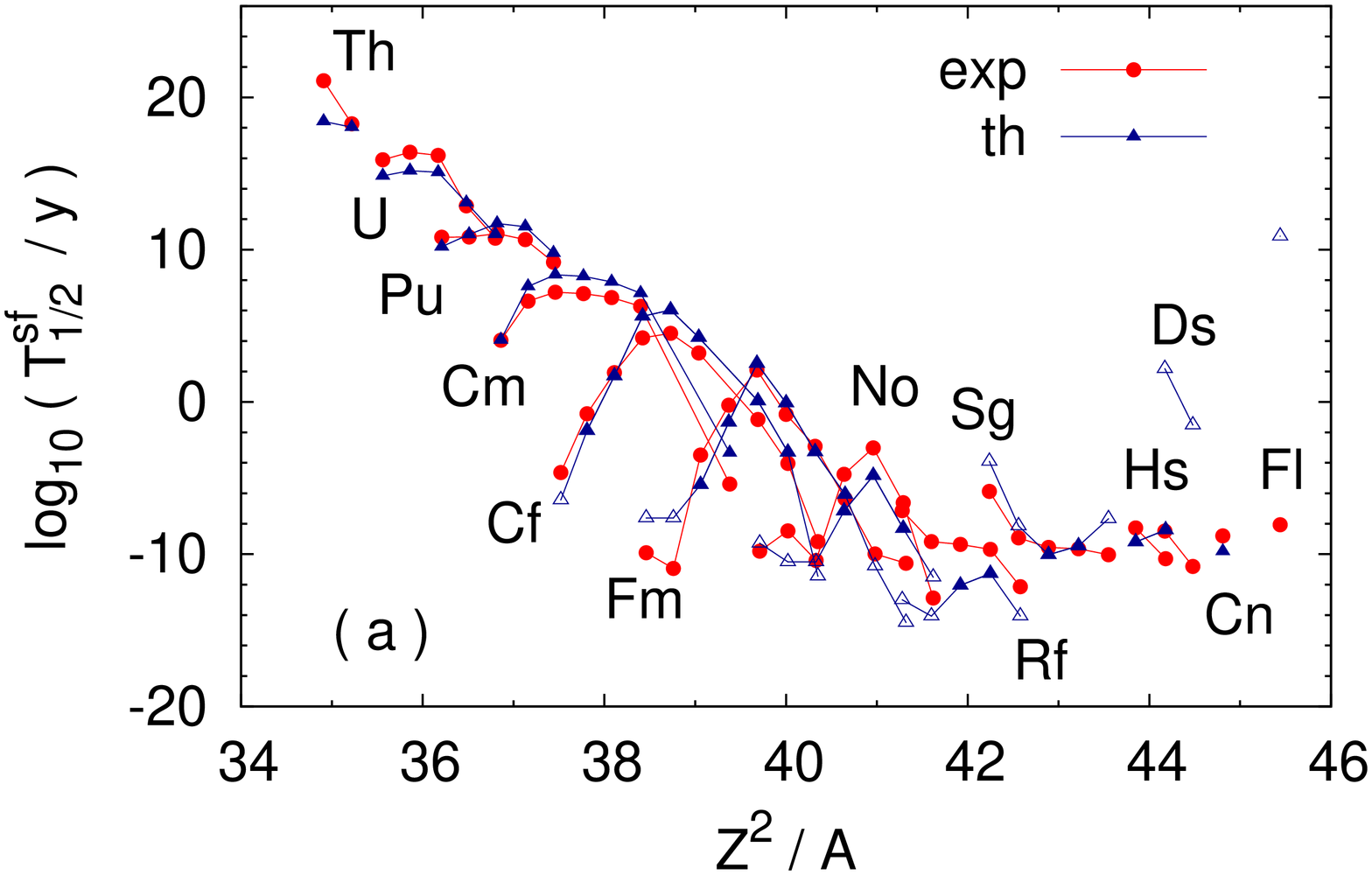}\\
\vskip-1.5cm
\includegraphics[width=0.83\columnwidth, angle=270]{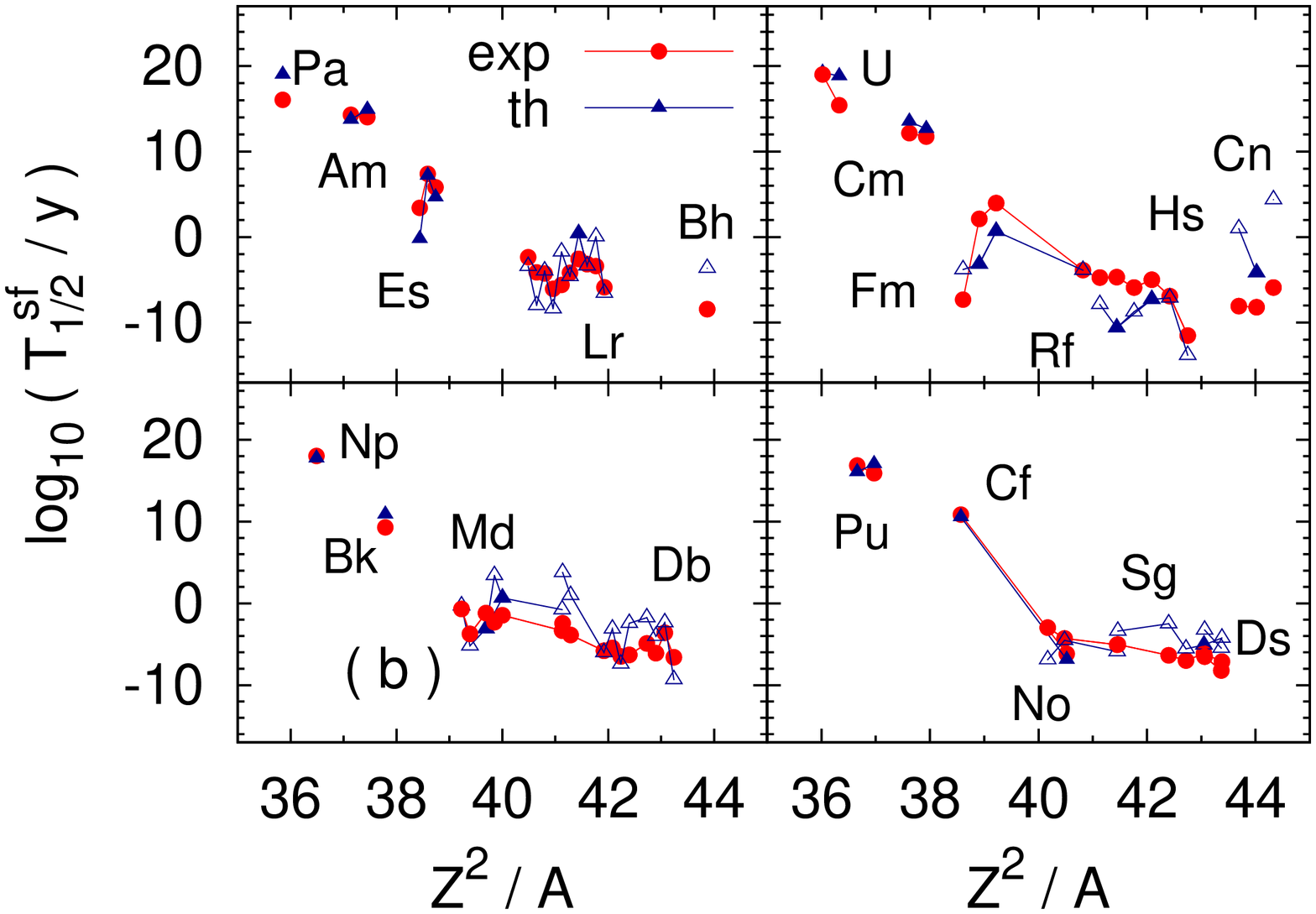}
\caption{Spontaneous fission half-lives of even-even (a) and odd (b) nuclei,
calculated using formula (\ref{Tsf}) (triangles) in comparison to the
experimental values \cite{nudat2} (dots). }
\label{fig_tsf}
\end{figure}

Spontaneous fission half-lives, calculated using formula Eq.~(\ref{Tsf}) are
presented and compared with the data in Fig.~\ref{fig_tsf}. Isotopes which
for masses are still not measured are marked with open symbols. A very
good accuracy was achieved in actinides region, where the root-mean-square
deviation of $\log T_{1/2}^{\rm sf}$ is equal to 1.18 and it grows to 1.57 when
odd-$A$ and odd-odd nuclei are taken into account. The r.m.s. deviation reaches
1.96, when one includes to the analysis the super-heavy isotopes, for which
the liquid drop barrier vanishes. The quality of reproduction of known
life-times by this simple model is even better than that presented in
Fig.~\ref{dynamic} obtained in the macroscopic-microscopic model with the
cranking inertia \cite{Ba81}. Also more advanced modern calculations based on
the HFB theory with the Skyrme (see e.g. \cite{SBN13,SMB13} or Gogny force
\cite{RGR14} did not bring better estimates for large sample of isotopes than
the simple formula presented above. It raises a question: why the simple
\'Swi\c{a}tecki model works? To understand it, one has to recall
\'Swi\c{a}tecki's topographical theorem and the results obtained using dynamic
(least action) trajectories to fission described in Section~\ref{SF}.

\subsection{The topographical theorem}
\begin{figure}[h!]
\begin{centering}
\includegraphics[width=\columnwidth]{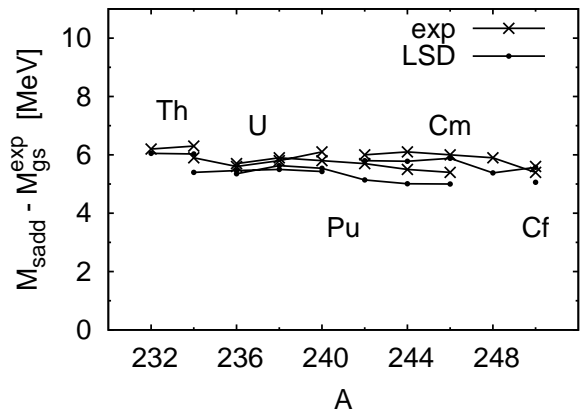}
\caption{Experimental fission barrier heights compared with their estimates
evaluated according to the topographical theorem of \'Swi\c{a}tecki and the LSD
macroscopic energy. After Ref.~\cite{DNP09}.}
\label{topo}
\end{centering}
\end{figure}

It was shown in Refs. \cite{PD03,DPB07,IP09} that the LSD model (\ref{LSD}),
which parameters were fitted to the experimental ground-state masses only, is
able to reproduce well the fission barrier heights of light, medium and heavy
nuclei, when the microscopic part of the ground-state binding energy is taken
into account. According to the topographical theorem, proposed by Myers and
\'Swi{}\c{a}tecki~\cite{MS96}, the mass of a nucleus in a saddle-point is mostly
determined by the macroscopic part of its binding energy. The influence of the
shell effects on the saddle-point energy is rather weak and in the first
approximation might be neglected. So, one can evaluate the fission barrier
height as a difference between the macroscopic (here LSD) and the experimental
mass:
\begin{equation}
 V_B(Z,N) = M_{\rm LSD}(Z,N,{\rm saddle}) - M_{\rm exp}^{g.s.},
\label{vb}
\end{equation}
where $M_{\rm LSD}(Z,N,{\rm saddle})$ is a macroscopic part of LSD mass formula
(\ref{LSD}) taken in a saddle point.

The comparison of the experimental fission barrier height of actinide nuclei
with their estimates, evaluated according to the topographical theorem of
\'Swi\c{a}tecki using the LSD macroscopic energy, is presented in Fig.~\ref{topo}.
The r.m.s. deviation of the estimates from the data for all 18 considered
fission barriers heights is 0.31 MeV only, what proves, that the topographical
theorem really works \cite{DNP09}. All estimates do not exceed 0.67 MeV and lie
below the experimental values (except $^{250}$Cf), what give a place for not
washed out shell effects around the saddle point energy.

\subsection{Justification of the simple formula for $T_{\rm sf}$}

Let us consider one dimensional fission barrier along the least-action
(dynamic) trajectory described in Sec.~\ref{SF}. Assuming, that the fission path
is parametrised by the collective coordinate $s$, one can evaluate the potential
$V(s)$ and the mass parameter $B_{ss}(s)$ corresponding to this path. The
fluctuations of the inertia $B_{ss}$ along the least-action trajectory becomes
smaller and smaller, when one increase the number of collective coordinates.
Especially the collective pairing degrees of freedom significantly wash out the
fluctuations of the inertia function \cite{SPP89}.
\begin{figure}[h]
\begin{centering}
\includegraphics[width=\columnwidth]{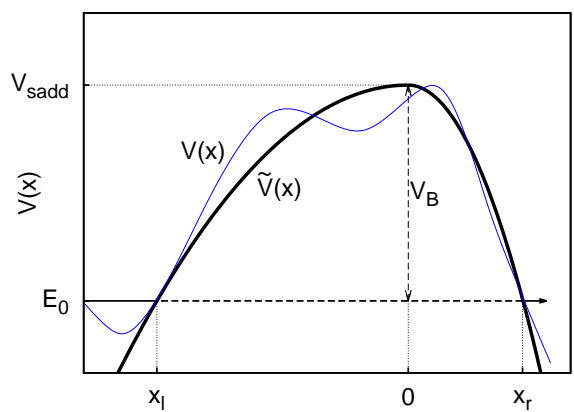}
\caption{Schematic plot of a fission barrier in form of two inverted
parabolas smoothly joined at the scission point.}
\label{abar}
\end{centering}
\end{figure}

A simple transformation:
\begin{equation}
x(s)=\int\limits_{s_{\rm sadd}}^s \sqrt{\frac{B_ss(s')}{m}} ds'\,\,,
\label{tran}
\end{equation}
bring us to the new coordinate $x(s)$, in which the inertia becomes constant
$B_{xx}=m$. The lower integration limit in Eq.~(\ref{tran}) is chosen at point $x=0$, corresponding to the saddle point $s_{\rm sadd}$. The collective
potential in the new coordinate, schematically plotted in Fig.~\ref{abar},
can be approximated by two inverted parabolas of different stiffnesses $C_l$
and $C_r$ having the maximum at the saddle point:
\begin{equation}
\widetilde V(x)= \left\{
\begin{array}{ll}
V_{\rm sadd} - \frac{1}{2}\,C_l\, x^2~~~&{\rm for}~~~x\,<\,0\,\,,\\[1ex]
V_{\rm sadd} - \frac{1}{2}\,C_r\, x^2~~~&{\rm for}~~~x\,>\,0\,\,,
\end{array}\right.
\end{equation}
The stiffnesses of the $\widetilde V$ potential are chosen in such a way that
the action-integral $S$ (\ref{act}) becomes equal:
\begin{equation}
\int\limits_{-x'_l}^{x'_r}\!\!\sqrt{\frac{2m}{\hbar^2}[V(x)-E_0]}\,dx =\!
 \!\int\limits_{-x_l}^{x_r}\!\!\sqrt{\frac{2m}{\hbar^2}[\widetilde
V(x)-E_0]}\,dx
\label{ss}
\end{equation}
where pairs ($-x'_l,x'_r$) and ($-x_l,x_r$) are classical left and right
turning points for the true and approximative potential respectively. The last
integral in Eq.~(\ref{ss}) can be rewritten as
$$
S=\!\sqrt{\frac{2m}{\hbar^2}}\left\{\!\int\limits_0^{x_l}\!\sqrt{V_B-\frac{
C_lx^2}{2}}\,dx
 +\!\!\int\limits_0^{x_r}\!\sqrt{V_B-\frac{C_rx^2}{2}}\,dx\right\},
$$
where $V_B=V_{\rm sadd}-E_0$ is the fission barrier height.\\
After a small algebra the action integral becomes
\begin{equation}
S=\frac{\pi}{2\hbar}V_B\left(\sqrt{\frac{m}{C_l}}+\sqrt{\frac{m}{C_r}}\right)
 =\frac{\pi}{\hbar}V_B\frac{\omega_l+\omega_r}{2\omega_l\omega_r} \,\,,
\end{equation}
where $\omega_l=\sqrt{C_l/m}$ and $\omega_r=\sqrt{C_r/m}$ are frequencies of
the left and right (inverted) oscillator respectively. Introducing the average
oscillator frequency
\begin{equation}
\widetilde\omega= \frac{2\omega_l\omega_r}{\omega_l+\omega_r} \,,
\end{equation}
one can bring the action integral to the following form:
\begin{equation}
S=\frac{2V_B}{\hbar\widetilde\omega}\,.
\label{sapp}
\end{equation}
In our approximation the action integral is proportional to the fission barrier
height measured in the energy quanta of the harmonic oscillator, which
approximates the fission barrier form. For the action-integral $S>1$ the
logarithm of the spontaneous fission half-lives (\ref{tsf}) can written as
\begin{equation}
\log(T^{\rm sf}_{1/2}/y)= 2S - \log(n) - 0.3665
\end{equation}
where the constant 0.3665=log[ln(2)] and $n$ is the number of assaults of
nucleus on the fission barrier per year ($y$). Having in mind, that according
to the topographical theorem the fission barrier height is $V_B=M_{\rm
LSD}^{\rm sadd}-M_{exp}^{gs}$ and making use of the relation (\ref{sapp}), one
can rewrite the last equation as follows:
\begin{equation}
\log(T^{\rm sf}_{1/2}/y)+\frac{4\delta M^{\rm exp}_{\rm
micr}}{\hbar\tilde\omega}=
\frac{4V_B^{\rm LSD}}{\hbar\tilde\omega} - \log(n) - 0.3665 ~,
\label{t12m}
\end{equation}
where $\delta M^{\rm exp}_{\rm micr}$ was defined in Eq.~(\ref{dM}) and
$V_B^{\rm LSD}=M_{\rm LSD}^{\rm sadd}-M_{\rm LSD}^{\rm sph}$ is the liquid
drop fission barrier height. The right hand side of Eq.~(\ref{t12m}) is a very
smooth function of nucleon numbers as it is defined only by global properties
of nucleus. Note, that the derived equation has the same structure as the
phenomenological formula of \'Swi\c{a}tecki (\ref{fZN}), what proves his
Ansatz.
\begin{figure}[h!]
\begin{centering}
\includegraphics[width=0.75\columnwidth, angle=270]{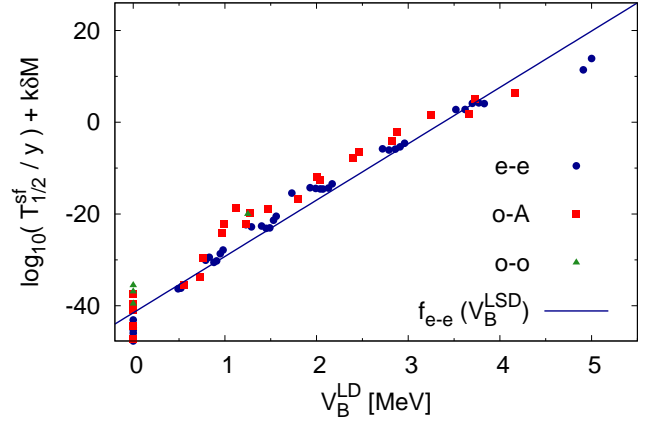}
\caption{Logarithms of the observed spontaneous fission half-lives corrected
with $7.7{\delta M}$ as a function of liquid drop barrier height.}
\label{vb_lsd}
\end{centering}
\end{figure}

The liquid drop barrier height of actinides decreases almost linearly in
function of $Z$ from 4.3 MeV for $Z=90$ to 0 for $Z\ge 103$ \cite{IP09}. The
fission barrier of finite height appears in the super-heavy nuclei mostly due
to the shell effects in the ground state. The smooth dependence of logarithms
of spontaneous fission half-lives, corrected by the ground state
shell-plus-pairing effects, on the LSD fission barrier heights, shown in
Fig.~\ref{vb_lsd} for even-even (e-e), odd A (o-A) and odd-odd (o-o) nuclei.
The data for e-e isotopes lie very close to the straight line, what validates
Eq.~(\ref{t12m}). The data for o-A and o-o are above this line, what is due to
the specialization energy, which increase the fission barrier heights. As it was mentioned before, stability of super-heavy nuclei is determined by shell effects. The liquid drop barrier of these isotopes vanishes, what is visible in Fig. \ref{vb_lsd} for nuclei with $Z\ge103$.

\section{Conclusions}

The following conclusions can be drawn from our investigation:\\[-4ex]
\begin{itemize}
\item Simple, consistent model was applied to reproduce alpha decay half-lives
in 360 nuclei with atomic number $52 \le$ Z $\le 110$.\\[-3ex]
\item Model reproduces $\alpha-$decay half-lives with quite good accuracy; the
root-mean-square deviation of $\rm log_{10}(T^{\alpha}_{1/2})$ for even-even
isotopes is equal to 0.39.\\[-3ex]
\item Large underestimations in half-lives of $N=127$ isotones arises from
strong shell effects, which are not considered in this simple model.\\[-3ex]
\item Semi-empirical formula for the spontaneous fission half-lives, depending
on proton number and the ground state microscopic corrections, reproduces data
for even-even super-heavy nuclei with reasonable accuracy.\\[-3ex]
\item Quality of spontaneous fission half-lives evaluation breaks down for
nuclei with not measured yet masses.\\[-3ex]
\item The logarithms of the spontaneous fission half-lives, corrected by the
ground state shell-plus-pairing effects, are roughly proportional to the
macroscopic barrier heights in nuclei up to Z=102.\\[-5ex]
\end{itemize}
\section{Acknowledgements}
This work was supported by the Polish National Science Centre grant
No.~2013/11/B/ST2/04087.\\[-4ex]

\end{document}